\documentclass[12pt]{iopart}
\usepackage{graphicx}

\begin{document}

%% NOTE: TITLE PAGE & TOC NOT USED FOR MANUSCRIPT SUBMISSIONS %%
%\title{Spectral decomposition of entangled photons with an arbitrary pump}

%\vskip4pc

%\tableofcontents
%\clearpage
%% NO TITLE PAGE FOR OPEX SUBMISSIONS %%

%% START HERE
%%%%%%%%%%%%%%%%%% title page information %%%%%%%%%%%%%%%%%%
\title[Spectral decomposition of entangled photons with an arbitrary pump]{Spectral decomposition of entangled photons with an arbitrary pump}

\author{Alison M. Yao}

\address{SUPA and Department of Physics, University of Strathclyde, Glasgow G4 0G, Scotland, U.K.}

\ead{alison@phys.strath.ac.uk} %% email address is required


\begin{abstract}
We calculate the bi-photon state generated by spontaneous parametric down conversion in a thin crystal and under 
collinear phase matching conditions using a pump consisting of any superposition of Laguerre-Gauss modes. 
The result has no restrictions on the angular or radial momenta or, in particular, on the width of the pump, signal and 
idler modes. We demonstrate the strong effect of the pump to signal/idler width ratio on the composition of the 
down-converted entangled fields. Knowledge of the pump to signal/idler width ratio is shown to be
essential when calculating the maximally entangled states that can be produced using pumps with
a complex spatial profile.
\end{abstract}

\maketitle

%%%%%%%%%%%%%%%%%%%%%%%%%%%%%%%%%%%%%%%%%%%%%%%%%%%%%%%%%%%%%%%%%
\section{Introduction}
Quantum entanglement is one of the defining properties of quantum mechanics. It forms the basis for quantum
information \cite{Plenio07,Barnett} and quantum computing \cite{Neilsen} and is essential for applications such as entangled 
cryptographic systems \cite{Ursin07} and quantum imaging \cite{DAngelo04,Bennink04,Pittman95,QIbook}. Much of the research in 
this area has focussed on qubit quantum entanglement, although entanglement has also been demonstrated between 
spatial modes carrying orbital angular momentum (OAM) \cite{Mair01,Jack10,Leach10}. Since these modes are defined 
within an infinite-dimensional, discrete Hilbert space there is considerable interest in their potential to generate multi-dimensional 
entangled states not only for enhancing the efficiency of current quantum protocols, but also for the realization of new, higher
dimensional quantum channels. 

Spontaneous parametric down conversion (SPDC), the generation of two lower-frequency photons when a pump field interacts 
with a nonlinear crystal, has been shown to be a reliable source of photons which are entangled \cite{Hong85} and, in
particular, in their OAM \cite{Mair01}. The spatial structure of the down-converted biphotons can be expressed as a superposition of 
Laguerre-Gauss modes of different amplitudes, with the width of the modal expansion relating to the amount of entanglement 
of the final state. In this paper, we calculate the exact
analytical form of the biphotons for any Laguerre-Gauss pump, with no restrictions on the angular or radial
momenta, $\ell$ and $p$, respectively, or, in particular, on the width, $w$, of the pump, signal and idler modes.
Our calculation demonstrates that OAM is conserved.

We use our result to calculate the coincidence amplitudes of the LG modes in the generated two-photon entangled state
for a variety of pump beams. We demonstrate the the importance of the various beam sizes (the pump width and the 
choice of LG base for the signal and idler) in determining the state of the down-converted photon and the resulting $\ell$ 
distribution, also known as the quantum spiral bandwidth (SB) \cite{Torres03}.
We find excellent agreement with previous analyses which considered more restricted conditions \cite{Torres03,Ren04}.

Knowledge of the exact form of the down-converted photons opens up the possibility of engineering the entangled modes, 
through an appropriate choice of pump beam, in order to produce biphotons which are, for example, maximally entangled in
some subspace \cite{Torres03b}. We show how our result can be easily extended to pump beams which are complex 
superpositions of Laguerre-Gauss modes, including pump modes containing phase singularities. We again see excellent agreement 
with previous analysis \cite{Torres03b}. Moreover, by allowing the beam widths to be free parameters, we show that
the pump to signal/idler size plays a critical role in the engineering of the down-converted state, with states only being
maximally entangled at particular width ratios.

%%%%%%%%%%%%%%%%%%%%%%%%%%%%%%%%%%%%%%%%%%%%%%%%%%%%%%%%%%%%%%%%%
\section{Spontaneous parametric down conversion and spiral bandwidth}
We consider a thin nonlinear crystal (typically $1$--$3$mm), tuned for type-I, collinear down-conversion, illuminated by a 
continuous-wave Laguerre-Gaussian pump beam propagating in the $z$-direction and confined within the crystal and 
assume perfect phase matching. 
This produces two highly-correlated, lower-frequency photons, commonly termed signal and idler. Since energy is conserved, 
$\omega_p = \omega_s + \omega_i$, where the subscripts $p, s, i$ refer to the pump, signal and idler, respectively.

The down-converted biphoton state in this case is given by \cite{Saleh00,Torres03b}
\begin{equation}
|\psi_{SPDC}\rangle = \int dr_{\perp} \Phi(r_{\perp}) \hat{a}_s^{\dagger}(r_{\perp})  \hat{a}_i^{\dagger}(r_{\perp}) |0,0\rangle
\end{equation}
where $\Phi(r_{\perp})$ is the spatial distribution of the pump beam at the input face of the crystal, $| 0,0 \rangle$ is the 
vacuum state and $\hat{a}_s^{\dagger}( r_{\perp})$, $\hat{a}_i^{\dagger}(r_{\perp})$ are creation operators for the signal 
and idler modes, respectively. The integral is over the plane perpendicular to the axis of the pump beam.

Since the photon pairs generated by SPDC are entangled in arbitrary superpositions of an infinite number of modes with 
OAM they are best described by mode functions that are Laguerre-Gauss modes, $LG_p^{\ell}$. At the beam waist ($z=0$) 
the normalised LG modes are given, in cylindrical co-ordinates, by \cite{Barnett07}:
\begin{eqnarray}
LG_p^{\ell} (\rho, \phi)=\sqrt{\frac{2 p!}{\pi \left( p+| \ell | \right)!}} \frac{1}{w} \left( \frac{\rho \sqrt{2}}{w} \right)^{| \ell |} 
 \exp \left( \frac{- \rho^2}{w^2} \right) L_p^{| \ell |}\left( \frac{2 \rho^2}{w^2} \right) \exp(i \ell \phi) .
\label{eqn:z0LGreal}
\end{eqnarray}
Here $w$ is the beam waist, $\ell$ corresponds to the angular momentum, $\ell \hbar$, carried by the beam and describes 
the helical structure of the wave front around a wave front singularity and $p+1$ describes the number of radial intensity maxima. \\

We can then write the biphoton state as:
\begin{eqnarray}
|\psi_{SPDC}\rangle = \sum_{\ell_s,p_s} \sum_{\ell_i,p_i} C_{p_s,p_i}^{\ell_s,\ell_i} \left| \ell_s, p_s; \ell_i, p_i \right \rangle \end{eqnarray}
where $P_{p_s,p_i}^{\ell_s,\ell_i}  = \left| C_{p_s,p_i}^{\ell_s,\ell_i} \right|^2 $
is the probability of finding one photon in the signal mode $\left| \ell_s, p_s \right \rangle$ and the other in the idler mode
$\left| \ell_i, p_i \right \rangle$ given a pump mode $\left| \ell_p, p_p \right \rangle$.

The coincidence amplitudes $C_{p_s,p_i}^{\ell_s,\ell_i}$ are calculated from the overlap integral
\begin{eqnarray}
C_{p_s,p_i}^{\ell_s,\ell_i} &=& \langle \psi_i,\psi_s|\psi_{SPDC} \rangle  \\
&=& \int_{0}^{2 \pi} \hspace{-0.2cm} d\phi \int_{0}^{\infty} \hspace{-0.2cm} \rho \, d\rho   
LG_{p_p}^{\ell_p}(\rho, \phi) \left [ LG_{p_s}^{l_s}(\rho, \phi) \right]^* \left [ LG_{p_i}^{l_i}(\rho, \phi) \right ]^* . 
\label{eqn:overlap}
\end{eqnarray}
Note that after substituting (\ref{eqn:z0LGreal}) into (\ref{eqn:overlap}) the integral over the azimuthal coordinate is
\begin{equation}
\int_{0}^{2 \pi} \hspace{-0.2cm} \, \, d \phi \exp \left [ i \left ( \ell_p -\ell_s - \ell_i \right ) \phi \right] = 2 \pi \delta_{\ell_p,\ell_s+\ell_i} ,
\label{eqn:consOAM}
\end{equation}
from which we obtain the well-known conservation law for the orbital angular momentum: $\ell_p = \ell_s + \ell_i$ \cite{Mair01, FrankeArnold02}.
Substituting this into (\ref{eqn:overlap}), and making use of the associated Laguerre polynomial in the form \cite{Gradshteyn}
\begin{eqnarray}
L_p^{|\ell|} (x) = \sum_{i=0}^p (-1)^i \frac{(p + |\ell|) !}{(p-i)! (|\ell|+i)! i!} \, x^i \, ,
\end{eqnarray}
we obtain
\begin{eqnarray}
C_{p_s,p_i}^{\ell_s,\ell_i} &=& \delta_{\ell_p,\ell_s+\ell_i} 
\sqrt{ \frac{2}{\pi w_p^2}} \,
\frac{2^{\sigma_{\ell} + 1} \gamma_s^{|\ell_s| + 1} \gamma_i^{|\ell_i| + 1}}{\left ( 1 + \gamma_s^2 + \gamma_i^2 \right )^{\sigma_{\ell} + 1}} \nonumber \\
&\times&\sqrt{p_p! p_s! p_i ! \left(| \ell_p | + p_p\right)! \left(| \ell_s | + p_s\right)! \left(| \ell_i | + p_i\right)!}  \nonumber \\
&\times&
\sum_{k=0}^{p_p} \sum_{i=0}^{p_s} \sum_{j=0}^{p_i}  
 \frac{(-2)^{k+i+j} \gamma_s^{2 i} \gamma_i^{2 j}}{\left ( 1 + \gamma_s^2 + \gamma_i^2 \right )^{k + i + j}} \nonumber \\
&\times& \frac{\left [ \sigma_{\ell} + k + i + j\right] ! }{(p_p-k)! (|\ell_p |+k)! k! (p_s-i)! (|\ell_s|+i)! i! (p_i-j)! (|\ell_i|+j)! j!} ,
\label{eqn:fullsol}
\end{eqnarray}
where $\sigma_{\ell} = \left ( |\ell_p| +  |\ell_s| + |\ell_i| \right ) /2$ and  
$\gamma_s = w_p / w_s$, $\gamma_i = w_p / w_i$ are the ratios of the pump width to the signal/idler widths. 

Equation (\ref{eqn:fullsol}) allows us to calculate the exact analytical form of the down-converted photons 
produced using any Laguerre-Gaussian pump (or superposition thereof) with no restrictions on any of the beam 
parameters. Calculation of the coincidence amplitudes can also be used to calculate the number 
of OAM modes participating in the down-converted state, otherwise known as the spiral bandwidth (SB).

%%%%%%%%%%%%%%%%%%%%%%%%%%%%%%%%%%%%%%%%%%%%%%%%%%%%%%%%%%%%%%%%%
\section{Coincidence amplitudes and spiral bandwidth for special cases}
For a given pump, equation (\ref{eqn:fullsol}) shows that the coincidence amplitudes have a strong dependence on the 
ratio of the pump width to the chosen widths of the LG bases for the signal and idler, $\gamma_s$ and $\gamma_i$. 
Experimentally, the maximum width of the pump is limited by the size of the down-conversion crystal, while the widths of 
the signal and idler beams are determined by the size of the single mode fibre used in the detection process. 
We can see this dependence more clearly if we consider some simplified examples.

\subsection{Gaussian pump, $\ell_p = p_p =0$.}
We first consider the simplest possible case: a Gaussian pump with $\ell_p = p_p =0$ and biphotons with no radial 
dependence i.e. $p_s = p_i = 0$. Moreover, we assume that the signal and idler modes have the same width s.t. 
$\gamma_s = \gamma_i = \gamma$. From from conservation of OAM, equation (\ref{eqn:consOAM}), we know 
that $\ell_s = -\ell_i = \ell$ in this case and so we can reduce equation (\ref{eqn:fullsol}) to
\begin{equation}
C_{0,0}^{\ell,-\ell} = 
\sqrt{ \frac{2}{\pi w_p^2}} \,
\left ( \frac{2 \gamma^2}{1 + 2 \gamma^2} \right )^{|\ell| + 1} .
\label{eqn:GPcoinc}
\end{equation}
It is clear that the SB depends only on the ratio of the pump to signal/idler widths, $\gamma$. 
As in previous work, for example \cite{Torres03,Ren04}, this further reduces to 
\begin{equation}
C_{0,0}^{\ell,-\ell} \propto \left ( \frac{2}{3} \right )^{|\ell|} 
\end{equation}
when all the beam widths are the same, i.e. $\gamma = 1$.

\begin{figure}[htbp] 
\centering
\includegraphics[width=8cm]{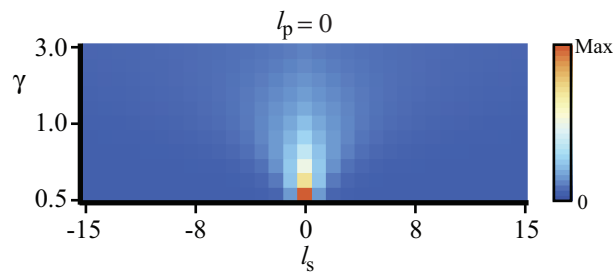} 
\caption{Spiral bandwidth for a Gaussian pump, $\ell_p = p_p = 0$ and fixed width $w_p = 1.0$ as a function
$\gamma$.} 
\label{fig:figure1}
\end{figure}

\begin{figure}[htbp] 
\centering
\includegraphics[width=\textwidth]{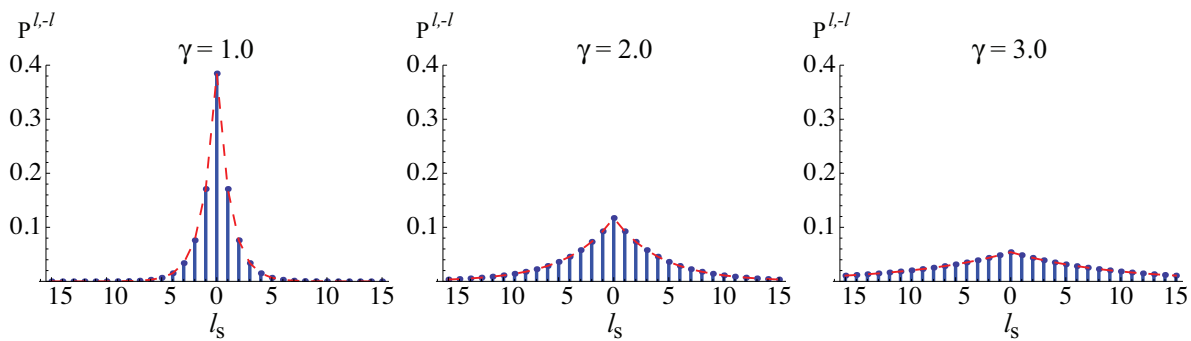} 
\caption{Spiral bandwidth for a Gaussian pump, $\ell_p = p_p = 0, w_p = 1.0$.  The signal
and idler widths are equal and reduced s.t. $\gamma = 1.0, 2.0, 3.0$ from left to right. 
Vertical (blue) lines are calculated using (\ref{eqn:GPcoinc}) and dashed (red) lines are calculated using eqn.s (10) 
\& (11) of \cite{Torres03}.} 
\label{fig:figure2}
\end{figure}

The effect of $\gamma$ on the spiral bandwidth is shown in figure (\ref{fig:figure1}) where we have
plotted the SB as a function of $\gamma$. It can be seen that as the signal/idler widths are reduced 
w.r.t. the pump width (i.e. as $\gamma$ increases) the SB increases but that the 
amplitudes of the participating modes decrease. In figure (\ref{fig:figure2}) we plot (in blue) the SBs for
$\gamma = 1.0, 2.0$ and $3.0$. These results are in excellent agreement with previous work, see, 
for example, figure (1) of \cite{Torres03} (shown in red in figure (\ref{fig:figure2})). Note that
throughout this work although we only plot the SB from $-15 \le \ell \le 15$, which are experimentally
realisable values, we calculate the coincidence amplitudes for a much larger range of $\ell$ and use 
these results to ``normalize'' the coincidence probabilities.
The effect of the different beam parameters with a Gaussian pump has recently been explored \cite{Miatto11}. 
Note that the results are also valid for non-collinear geometries in the approximation of a thin crystal. 

%------------------------------------------------------------------------------------------------------------------------------------------------------
\subsection{Laguerre-Gaussian pump, $\ell_p >0$.}
We now consider what happens when the pump is a higher-order Laguerre-Gaussian 
mode, i.e. $\ell_p > 0$. In this case conservation of OAM requires $\ell_i = \ell_p - \ell_s$. As before, we start with the simplest 
case: $p_{p,s,i} = 0$ and assume that the signal and idler modes have the same width. The probability amplitudes are then described by
\begin{eqnarray}
&&C_{0,0}^{\ell_s,\ell_i} = \delta_{\ell_p,\ell_s+\ell_i} 
\sqrt{ \frac{2}{\pi w_p^2}}  \frac{2^{\sigma_{\ell} + 1}  \sigma_{\ell} ! }{\sqrt{|\ell_p |! |\ell_s|! |\ell_i|!}} 
\frac{\gamma_s^{|\ell_s| + 1} \gamma_i^{|\ell_i| + 1}}{\left ( 1 + \gamma_s^2 + \gamma_i^2 \right )^{\sigma_{\ell} + 1}}  \nonumber
\end{eqnarray}
where $\sigma_{\ell} = \left ( |\ell_p| +  |\ell_s| + |\ell_i| \right ) /2$. This agrees with equation (8) of \cite{Ren04} when
using normalized LG modes and assuming that all of the modes have the same width.

\begin{figure}[htbp]
\centering
\includegraphics[width=\textwidth]{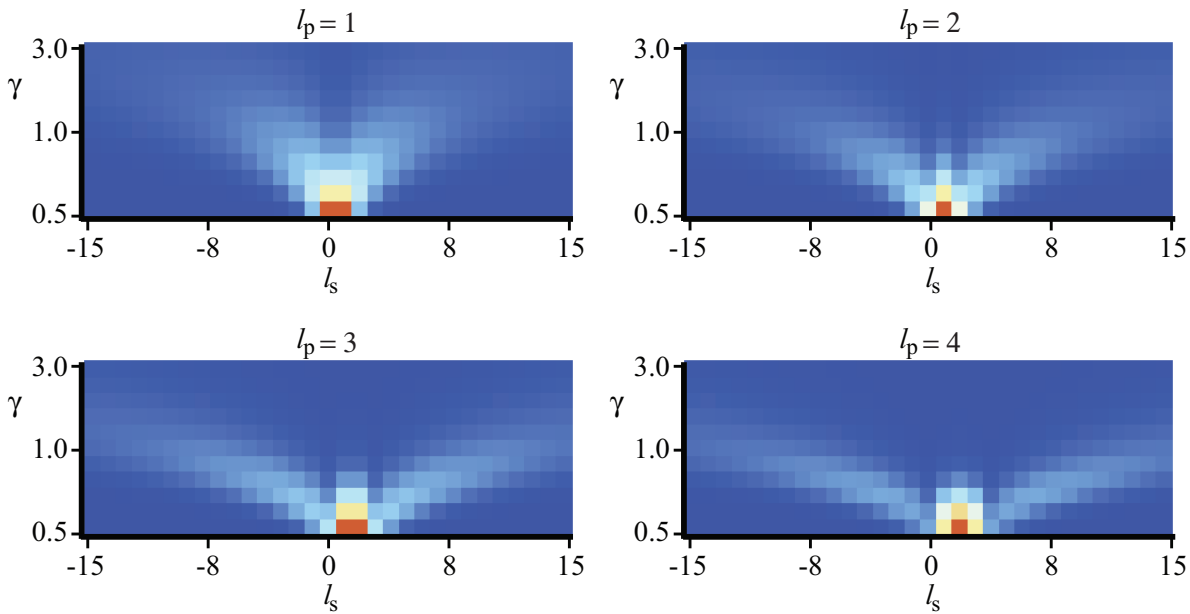} 
\caption{Spiral bandwidth for a Laguerre-Gaussian pumps of fixed width $w_p = 1.0$ and $\ell_p = 1, 2, 3, 4$ as the signal/idler 
width is reduced s.t. $\gamma_s = \gamma_i$ goes from $0.5$ to $3.0$.}
\label{fig:figure3}
\end{figure}

To see how the OAM of the pump affects the spiral bandwidth we first calculate the coincidence probabilities, 
$P_{0,0}^{\ell_s,\ell_p-\ell_s}$, as a function of $\ell_s$ and $\gamma$ for different values of $\ell_p$. 
As figure (\ref{fig:figure3}) shows, the SB is symmetric around the OAM value of the pump.
For small values of $\gamma$ the SB has a maximum at $\ell_s = \ell_p$.
As $\gamma$ is increased, however, the SB splits into two ``wings'' for $\ell_p > 0$, with a minimum at
$\ell_s = \ell_p$, and the amplitude increasing with $\ell_s$. The physical explanation of this is that pump
modes with $\ell_p > 0$ form a rings whose diameter increases with both the value of $\ell_p$ and the width of
the beam. Maximum overlap between the pump and the signal/idler modes, and hence maximum coincidence
amplitudes, therefore occurs for larger values of $\ell_s,\ell_i$.

\begin{figure}[htbp]
\centering
\includegraphics[width=\textwidth]{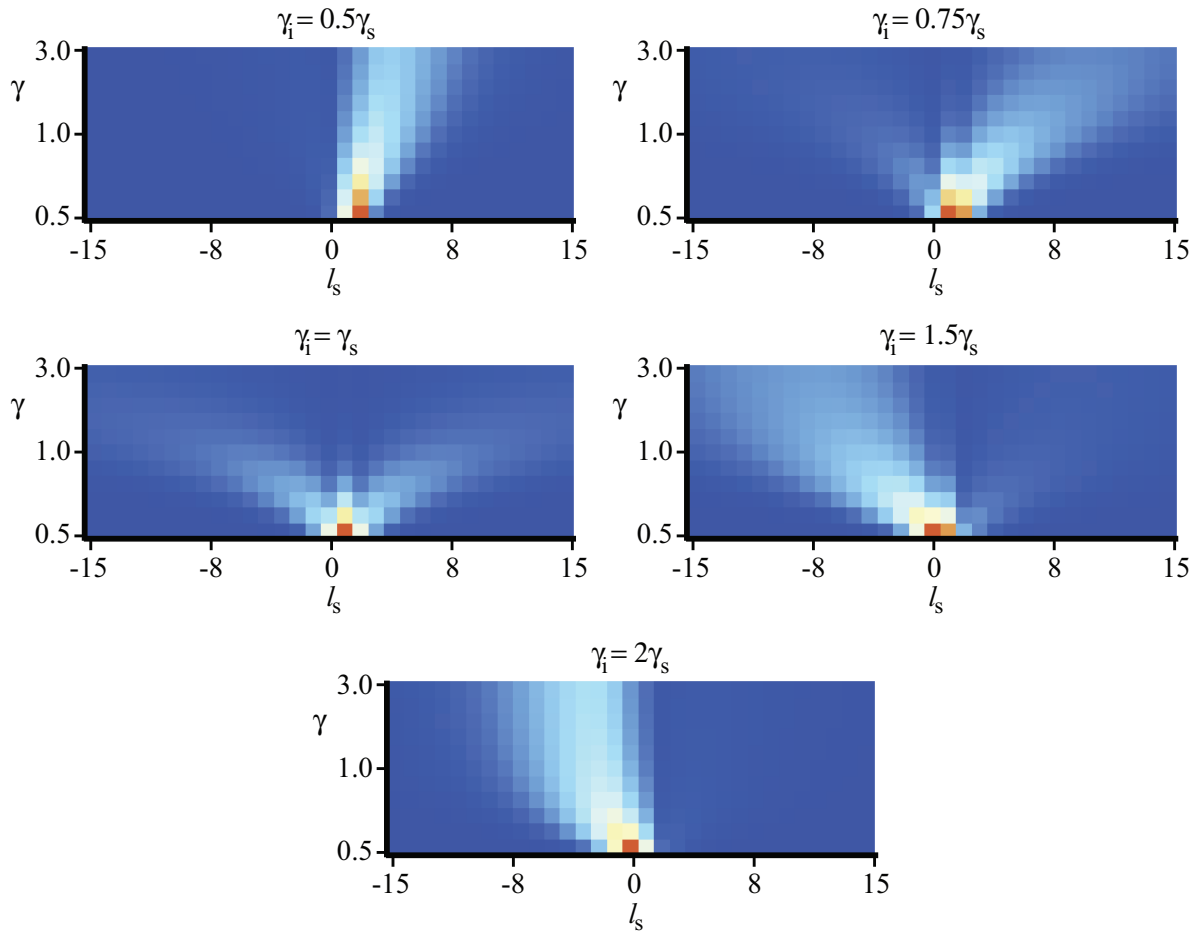} 
\caption{Spiral bandwidth for a Laguerre-Gaussian pump with $\ell_p = 2, w_p = 1.0$  as the signal/idler 
width is reduced s.t. $\gamma_s$ goes from $0.5$ to $3.0$. The signal and idler modes have different width ratios  
$\gamma_i/\gamma_s = 0.5, 0.75,1.0,1.5, 2.0$.}
\label{fig:figure4}
\end{figure}

We next consider what happens if the widths of the signal and idler are different. We find that by varying the 
ratio of the signal to idler widths (in this case from $\gamma_i = 0.5\gamma_s$ to $2.0 \gamma_s$) it is possible
to fully suppress the SB for either negative or positive values of $\ell_s$, as shown in figure (\ref{fig:figure4}).

\begin{figure}[htbp]
\centering
\includegraphics[width=\textwidth]{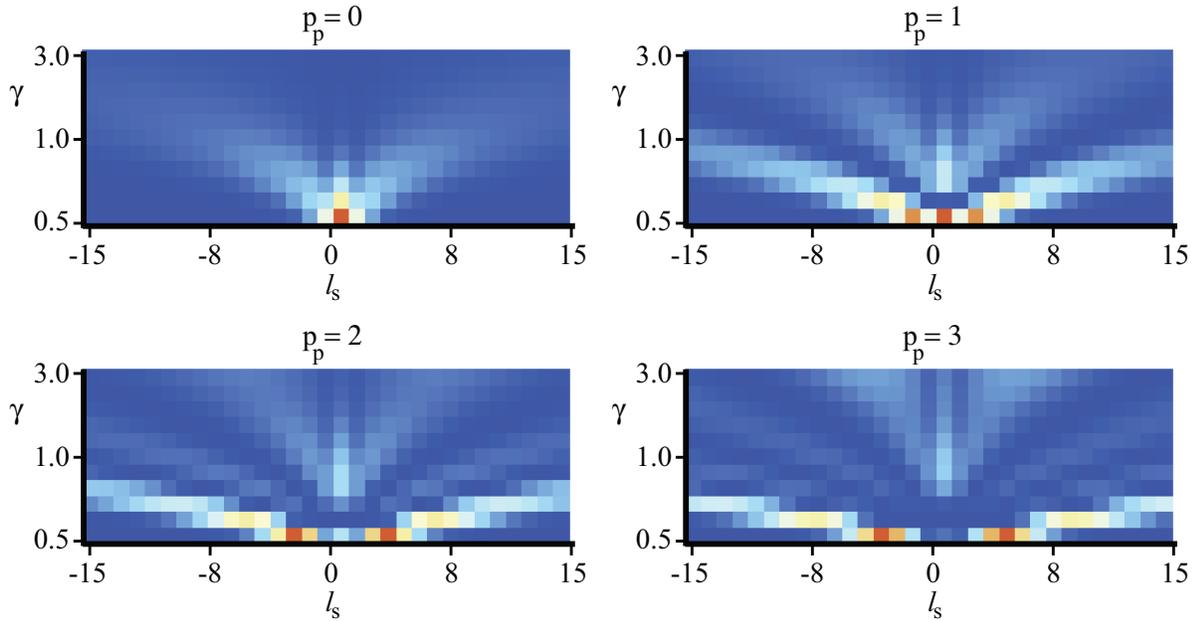} 
\caption{Spiral bandwidth for a Laguerre-Gaussian pumps with $\ell_p = 2, w_p = 1.0$ as the signal/idler 
width is reduced s.t. $\gamma_s = \gamma_i$ goes from $0.5$ to $3.0$. The radial number of the pump is increased
from $p_p = 0$ to $3$.}
\label{fig:figure5}
\end{figure}

Until now we have considered only modes with no radial modes, i.e. $p_p = p_s = p_i = 0$. Higher-order
LG beams are commonly produced using a spatial light modulator (SLM). Since these produce OAM modes 
with a range of radial indices, $p$, \cite{Arlt98}, we now look at the effect $p$ has on the form of the spatial 
mode function of the entangled photons. 
We find that the number of ``wings'' increases
with the radial index of the pump as shown in figure (\ref{fig:figure5}). As before, altering the ratio of the 
signal to idler width suppresses one set of wings. Again these correspond to areas of maximum overlap between 
the pump and the signal/idler modes.

\begin{figure}[htbp]
\centering
\includegraphics[width=\textwidth]{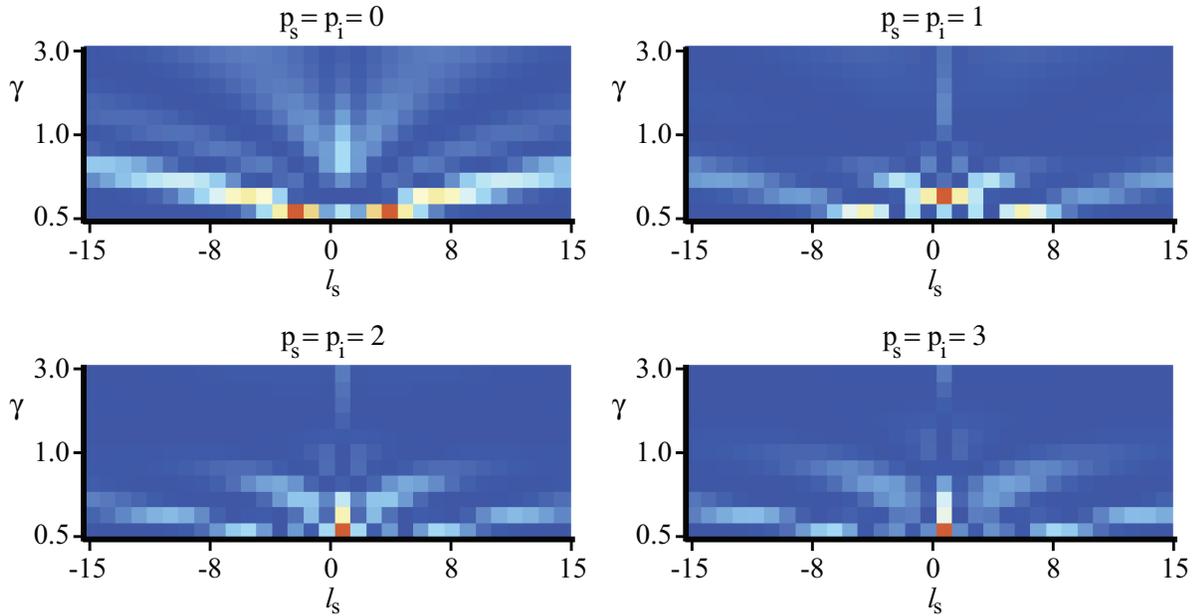} 
\caption{Spiral bandwidth for a Laguerre-Gaussian pumps with $\ell_p = p_p = 2, w_p = 1.0$ as the signal/idler 
width is reduced s.t. $\gamma_s = \gamma_i$ goes from $0.5$ to $3.0$. The radial number of the signal and idler modes is increased
from $p_s = p_i = 0$ to $3$.}
\label{fig:figure6}
\end{figure}

If we allow first our signal/idler beams to have non-zero radial modes and then all beams to have non-zero radial
modes the spiral bandwidth as a function of $\gamma$ becomes ever more complex. The effect of the radial number
of the pump is shown in figure (\ref{fig:figure6}).

%%%%%%%%%%%%%%%%%%%%%%%%%%%%%%%%%%%%%%%%%%%%%%%%%%%%%%%%%%%%%%%%%
\subsection{Engineered entangled states}

For many quantum communication applications entangling systems in higher-dimensional states is important. Not 
only does the higher dimensionality imply a greater potential for applications in quantum information processing, 
but it has also been suggested that increasing the dimensionality of the entangled states of a system can make 
its non-classical correlations more robust to the presence of noise and other detrimental effects 
\cite{Kaszlikowski00,Collins02}. 
Torres \textit{et al.} demonstrated (theoretically) that arbitrary engineered entangled states in any $d$-dimensional Hilbert 
space can be prepared using SPDC to translate the topological information contained in a pump beam into the amplitudes 
of the generated entangled quantum states \cite{Torres03b}. 
Controlling OAM state superpositions in this way engenders the ability to produce and manipulate quantum states with 
an arbitrarily large number of dimensions. 
Indeed, Torres \textit{et al.} illustrated this by calculating a number of maximally entangled states of different dimensions,
produced using pump beams containing phase singularities (PS).
Here we describe how our results can be easily extended to such pumps, which can be described by complex 
superpositions of LG modes \cite{Brambilla91}. Moreover, we demonstrate that the ratio of the pump to signal/idler 
widths must be taken into consideration in the calculation as it plays a critical role in the 
composition of the down-converted state: a state is generally maximally entangled for only one 
pump to signal/idler width ratio.

We first calculate the coincidence amplitudes resulting from a pump that is a superposition of Laguerre-Gauss 
modes of equal width, $w_p$, and complex amplitudes $a_n$. We can write this as 
\begin{equation}
\sum_n a_n LG_{p_n}^{\ell_n} (\rho, \phi) = a_1 LG_{p_1}^{\ell_1} (\rho, \phi) + a_2 LG_{p_2}^{\ell_2} (\rho, \phi ) + \ldots
\end{equation}
where $n$ labels the modes in the superposition and $\sum_n |a_n|^2 = 1$. The overlap integral (\ref{eqn:overlap}) 
then becomes
\begin{eqnarray}
C_{p_s,p_i}^{\ell_s,\ell_i} &=& \sum_n  \int_{0}^{2 \pi} \hspace{-0.2cm} d\phi \int_{0}^{\infty} \hspace{-0.2cm} \rho \, d\rho  
a_n LG_{p_n}^{\ell_n} (\rho, \phi) \left [ LG_{p_s}^{l_s}(\rho, \phi) \right]^* \left [ LG_{p_i}^{l_i}(\rho, \phi) \right]^* \nonumber \\
&=& \sum_n a_n C_{n,p_s,p_i}^{\; \; \; \, \ell_s, \ell_i},
\label{eqn:overlap2}
\end{eqnarray}
where $C_{n,p_s,p_i}^{\; \; \; \, \ell_s, \ell_i}$ are the coincidence amplitudes, calculated from (\ref{eqn:fullsol}), for each 
Laguerre-Gaussian component, $\left| \ell_p, p_p \right \rangle$, of the pump field.

In order to investigate a pump containing PSs we make use of its projection onto LG modes \cite{Indebetouw93}.
A field containing $N$ PSs can be written as a superposition of LG modes, 
$\sum_{\ell=0}^N a_{\ell} LG_0^{\ell}$, with complex amplitudes 
\begin{equation}
a_{\ell} = \sqrt{\pi} (-1)^{N-\ell} \left ( \frac{w_p}{\sqrt{2}} \right )^{\ell-1} \sqrt{\ell!}  \, \, b_{N-\ell}
\label{eqn:LGamps}
\end{equation}
where $b_n$ are given by equation (11) of \cite{Torres03b} or, equivalently, by
\begin{equation}
b_n = \frac{1}{m!} \frac{\partial^m}{\partial x^m} \prod_{i=1}^N \left ( 1 + \rho_i e^{i \phi_i} x \right ) \big|_{x=0}
\end{equation}
with $m = [0, N]$ and $\rho_i, \phi_i$ being the radial and azimuthal positions of the PSs, respectively.

\begin{figure}[htbp]
\centering
\includegraphics[width=12cm]{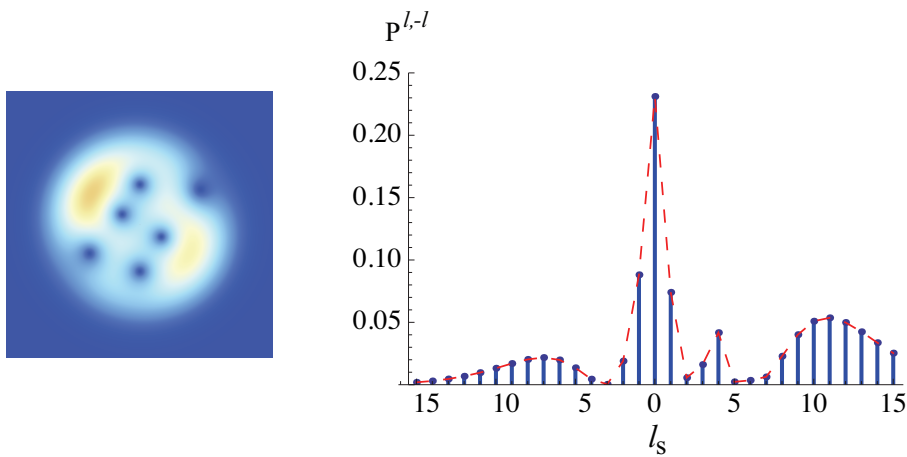} 
\caption{Pump containing six phase singularities, as described in \cite{Torres03b}, and resulting probability amplitudes. The blue 
lines show the results found using equation (14) of \cite{Torres03b} while the red line is the result found using 
equations (\ref{eqn:fullsol}) and (\ref{eqn:overlap2}).}
\label{fig:figure7}
\end{figure}

\begin{figure}[htbp]
\centering
\includegraphics[width=\textwidth]{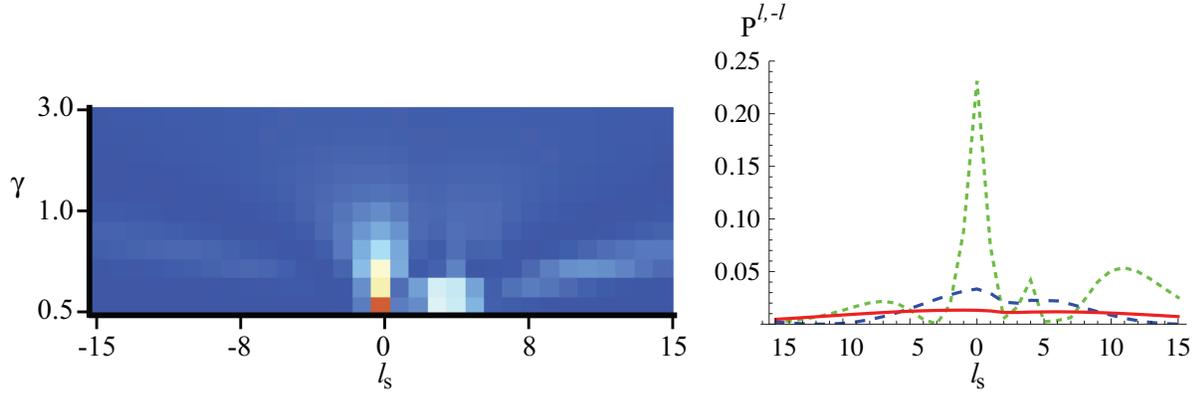} 
\caption{Spiral bandwidth for superposition pump for $\gamma_s = \gamma_i = 0.5$ to $3.0$ (left) with 
slices at $\gamma_s = \gamma_i = 1.0$ (green, small dash), $2.0$ (blue, large dash) and $3.0$ (red, solid) shown on right.} 
\label{fig:figure8}
\end{figure}

Following \cite{Torres03b}, we consider a pump containing six PSs at positions $\rho_1 =  0.65 w_p, \, 
\rho_2 = 1.85 w_p, \, \rho_3 = 1.06 w_p, \, \rho_4 = 0.54 w_p, \, \rho_5 = 1.53 w_p, \, \rho_6 = 1.24 w_p$ 
and $\phi_i = i \pi /3$ for $i = 1,6 $. 
Substituting these values into (\ref{eqn:LGamps}) and then into (\ref{eqn:overlap2}) allows us to calculate the 
form of the pump, as shown on the left in figure (\ref{fig:figure7}), and the resultant coincidence 
probabilities, vertical (blue) lines shown on the right in figure (\ref{fig:figure7}). We find excellent agreement with the results from eqn. (14) of 
\cite{Torres03b} (dashed red line). We calculated the coincidence probabilities $P_{0,0}^{0,0}, P_{0,0}^{1,1}, P_{0,0}^{2,2}, 
P_{0,0}^{3,3}$ and confirmed that these were equal and so this pump produces the maximally entangled qu-quart in the 
subspace $S_4 = \lbrace \left|0,0 \right>, \left|1,1 \right>, \left|2,2 \right>, \left|3,3 \right>\rbrace$.
Note, however, that in this example we assumed that the pump, signal and idler all had the same width i.e. 
$\gamma_s = \gamma_i = 1$. Since we have seen that the composition of the down-converted state is 
strongly dependent on the ratio of the pump to signal and idler widths we extended these results by calculating 
the spiral bandwidth as a function of $\gamma$, keeping  $\gamma_s = \gamma_i$ for 
simplicity. As expected, we see a large variation in SB for different $\gamma$, as shown on the left in figure (\ref{fig:figure8}). This 
can been seen more clearly on the right of figure (\ref{fig:figure8}), where we plot the SB at $\gamma = 1.0, 2.0, 3.0$. 
It is clear that by choosing the mode widths appropriately it is possible to tailor the spiral bandwidth in very different ways.

\begin{figure}[htbp]
\centering
\includegraphics[width=6cm]{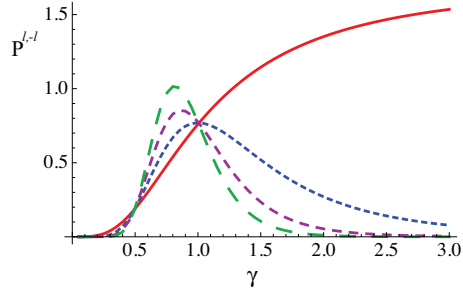} 
\caption{Coincidence probabilities $P_{0,0}^{0,0}$ (red, solid), $P_{0,0}^{1,1}$ (blue, small dash), $P_{0,0}^{2,2}$ (magenta, medium dash), 
and $P_{0,0}^{3,3}$ (green, large dash) as a function of $\gamma$. These are equal, and so the state is maximally entangled, 
for only 2 values of $\gamma$: $\gamma = 0.5, 1.0$.} 
\label{fig:figure9}
\end{figure}

Changing the relative beam sizes has a large effect on the composition of the entangled state
with important implications in the ``engineering'' of the down-converted state: states which are 
maximally entangled for one value of $\gamma$ are generally not maximally entangled for any other values
of $\gamma$ values. To illustrate this we consider the maximally entangled qu-quart produced by the six-PS 
pump above and calculate the coincidence probabilities $P_{0,0}^{0,0}, P_{0,0}^{1,1}, P_{0,0}^{2,2}, P_{0,0}^{3,3}$ 
as a function of $\gamma$. From figure (\ref{fig:figure9}) it is clear that this state is \textit{only} maximally entangled if 
$\gamma = 0.5$ or $\gamma = 1.0$. In order to obtain this state then, one has to ensure that these conditions
are met. 

%%%%%%%%%%%%%%%%%%%%%%%%%%%%%%%%%%%%%%%%%%%%%%%%%%%%%%%%%%%%%%%%%
\section{Conclusion}
We have calculated the exact analytical form of the biphotons produced from SPDC in a thin crystal and under 
collinear phase matching conditions. Our result has no restrictions on the angular or radial momenta or, 
in particular, on the width of the pump, signal and idler modes and can be used with any pump that is a 
Laguerre-Gauss mode, or superposition thereof. We have shown that OAM is conserved and found excellent 
agreement with previous analyses which considered more restricted conditions. 

We have used our result to calculate the spiral bandwidth for a variety of pumps and demonstrated the importance 
of the various beam parameters, in particular the pump width and the choice of LG base for the signal and idler, on 
the resulting coincidence amplitudes. 

In particular, we have demonstrated the importance of the beam widths when calculating engineered entangled
states and shown that maximal entanglement is dependent both on the form of the pump and on the ratio of the
beam widths. We suggest that this is an important consideration and may also offer another degree of freedom
in quantum communications.

%%%%%%%%%%%%%%%%%%%%%%%%%%%%%%%%%%%%%%%%%%%%%%%%%%%%%%%%%%%%%%%%%
\ack
This research was supported by the DARPA InPho program through the US Army Research Office award
W911NF-10-1-0395 and the UK Engineering and Physical Sciences Research Council (EPSRC). I would
particularly like to thank Stephen Barnett, Gian-Luca Oppo and Shashank Virmani for useful discussions.

% 

%%%%%%%%%%%%%%%%%%%%%%%%%%%%%%%%%%%%%%%%
% BIBLIOGRAPHY
%%%%%%%%%%%%%%%%%%%%%%%%%%%%%%%%%%%%%%%%


\section*{References}
\begin{thebibliography}{99}

\bibitem{Plenio07}
Plenio M B and Virmani S 2007 
\textit{Quantum Inf. Comput.} \textbf{7}, 1.

\bibitem{Barnett}
Barnett S M 2009 
\textit{Quantum Information} (Oxford University, New York).

\bibitem{Neilsen}
Neilsen M and Chuang I L 2000 
\textit{Quantum Computation and Quantum Information} 
(Cambridge University, Cambridge, England).

\bibitem{Ursin07}
Ursin R \textit{et al}. 2007 
\textit{Nature Phys.} \textbf{3}, 481. 

\bibitem{DAngelo04}
D'Angelo M, Kim Y H, Kulik S P and Shih Y 2004 
\PRL \textbf{92}, 233601. 

\bibitem{Bennink04}
Bennink R S, Bentley S J, Boyd R W and Howell J C 2004 
\PRL \textbf{92}, 033601.

\bibitem{Pittman95}
Pittman T B, Shih Y H, Strekalov D V and Sergienko A V 1995 
\PR A \textbf{52}, R3429.

\bibitem{QIbook}
\textit{Quantum Imaging} 2007  
Kolobov M I, ed.,
Springer, Singapore.

\bibitem{Mair01}
Mair A, Vaziri A, Weihs G and Zeilinger A 2001 
\textit{Nature} \textbf{412}, 313.

\bibitem{Jack10}
Jack B, Yao A M, Leach J, Romero J, Franke-Arnold S, Ireland D G, Barnett S M and Padgett M J 2010 
\PR A \textbf{81}, 043844.

\bibitem{Leach10}
Leach J, Jack B, Romero J, Jha A K, Yao A M, Franke-Arnold S, Ireland D G, Boyd R W, Barnett S M and Padgett M J 2010 
Science \textbf{329}, 662.

\bibitem{Hong85}
Hong C K and Mandel L 1985  
\PR A \textbf{31}, 2409.

\bibitem{Torres03} 
Torres J P, Alexandrescu A and Torner L 2003 
\PR A \textbf{68}, 050301.

\bibitem{Ren04}
Ren X F, Guo G P, Yu B, Li J and Guo G C 2004 
\JOB \textbf{6}, 243.

\bibitem{Torres03b} 
Torres J P, Deyanova Y and Torner L 2003  
\PR A \textbf{67}, 052313.

\bibitem{Saleh00}
Saleh B E A, Abouraddy A F, Sergienko A V and Teich M C 2000 
\PR A \textbf{62}, 043816.

\bibitem{Barnett07}
Barnett S M and Zambrini R 2007 
\textit{Quantum Imaging}, pg. 284, K. I. Kolobov, ed.,
Springer, Singapore.

\bibitem{FrankeArnold02}
Franke-Arnold S, Barnett S M, Padgett M J and Allen L 2002 
\PR A, \textbf{65}, 033823.

\bibitem{Gradshteyn}
Gradshteyn I S and Ryzhik I M 1980 
\textit{Table of Integrals, Series, and Products}, 8.970 (1),
Academic Press, Inc. (London) Ltd.

\bibitem{Miatto11}
Miatto F M, Yao A M and Barnett S M 2010 
Full characterisation of the quantum spiral bandwidth of entangled biphotons
\textit{Preprint} \PR A

\bibitem{Arlt98}
Arlt J, Dholakia K, Allen L and Padgett M J 1998 
\textit{J. Mod. Opt.} \textbf{45}, 1231.

\bibitem{Kaszlikowski00}
Kaszlikowski D, Gnacinski P, Zukowski M, Miklaszewski W and Zeilinger A 2000 
\PRL 85, 4418. 

\bibitem{Collins02}
Collins D, Gisin N, Linden N, Massar S and Popescu S 2002 
\PRL 88, 040404.

\bibitem{Brambilla91}
Brambilla M, Battipede F, Lugiato L A, Penna V, Prati F, Tamm C and Weiss C O 1991 
\PR A, \textbf{43}, 5090.  

\bibitem{Indebetouw93}
Indebetouw G 1993 
\textit{J. Mod. Opt.} \textbf{65}, 73.

\end{thebibliography}
\end{document}